\documentclass[twocolumn,english]{revtex4-1}
\usepackage[T1]{fontenc}
\usepackage[latin9]{inputenc}
\usepackage{geometry}
\usepackage{amsmath}
\usepackage{amssymb}
\usepackage{graphicx}
\usepackage{esint}

\makeatletter
\usepackage{braket}

\makeatother

\usepackage{babel}
\begin{document}
\title{Suppression of one-dimensional weak localization by band asymmetry}
\author{Kartikeya Arora}
\affiliation{Department of Physics, Indian Institute of Technology (Banaras Hindu
University), Varanasi 221005, India}
\author{Rajeev Singh}
\affiliation{Department of Physics, Indian Institute of Technology (Banaras Hindu
University), Varanasi 221005, India}
\author{Pavan Hosur}
\affiliation{Department of Physics and Texas Center for Superconductivity, University
of Houston, Houston 77004, USA}
\begin{abstract}
We investigate disorder-induced localization in metals that break
time-reversal and inversion symmetries through their energy dispersion,
$\epsilon_{k}\neq\epsilon_{-k}$, but lack Berry phases. In the perturbative
regime of disorder, we show that weak localization is suppressed due
to a mismatch of the Fermi velocities of left and right movers. To
substantiate this analytical result, we perform quench numerics on
chains shorter than the Anderson localization length $\xi$ -- the
latter computed and verified to be finite using the recursive Green's
function method -- and find a sharp rise in the saturation value
of the participation ratio due to band asymmetry, indicating a tendency
to delocalize. Interestingly, for weak disorder strength $\eta$,
we see a better fit to the scaling behavior $\xi\propto1/\eta^{2}$
for asymmetric bands than conventional symmetric ones.
\end{abstract}
\maketitle

\section{Introduction}

Weak localization (WL) refers to an enhanced tendency of electrons
in a disordered potential to localize due to constructive quantum
interference between pairs of time-reversed paths \citep{Localization_Theory_Experiment,Abrahams1979,WL_Anderson,WL_Thin_Film,WL_2D,Book_WL}.
It serves as a precursor to Anderson or \emph{strong }localization
which refers to the true arrest of quantum diffusion of free electrons
on a lattice \citep{Anderson58}. Since its conception in the context
of electrons in a metal, Anderson localization has been seen in light
waves \citep{Anderson_Light,Anderson_Light_2,Anderson_Light_Review},
ultrasound \citep{Anderson_Ultrasound,Anderson_Ultrasound_2}, ultracold
atoms \citep{Anderson_ultracold,Anderson_ultracold_3d,Anderson_ultracold_experiment,Anderson_ultracold_review},
and more recently, digital quantum simulators such as those provided
by IBMQ \citep{Anderson_digital,Anderson_50_Years}. 

The two regimes of localization are elegantly captured by the scaling
theory of localization \citep{Localization_Theory_Experiment,Anderson_Review_Lee_Ramakrishnan,Anderson_Transitions,Abrahams1979}.
Formulated as a renormalization group approach, it describes the scaling
behavior of conductance in disordered systems via a scaling function
that, in its simplest form, depends only on the conductance but has
no explicit dependence on system size. Then, the physical conductance
in the thermodynamic limit is given by the appropriate stable fixed
point of the renormalization group flow. The results of this procedure
depend sensitively on the dimensionality and symmetry class of the
system \citep{Localization_Theory_Experiment,Anderson_Review_Lee_Ramakrishnan,Anderson_Transitions,AltlandZirnbauer97,Wigner_1951,Dyson_1962}.
In three dimensions (3D), the scaling function has a zero of order
unity in every symmetry class, which is an unstable fixed point that
separates the localized and delocalized regimes at strong and weak
disorder, respectively. The localized phase corresponds to Anderson's
original prediction of localization in a disordered lattice. In 2D,
the asymptotic behavior of the system depends delicately on its symmetries.
If time-reversal symmetry ($\mathcal{T}$) is absent (unitary class)
or present but bosonic ($\mathcal{T}^{2}=1$, orthogonal class), disorder
is marginally relevant \citep{Localization_Theory_Experiment,Anderson_Review_Lee_Ramakrishnan,Anderson_Transitions,Hikami1980}.
Physically, this means infinitesimal disorder will eventually localize
the system in the thermodynamic limit, but the localization length
in practice can be astronomically large. This leads to striking experimental
signatures such as a sharp, symmetric cusp in the magnetoconductance
as the constructive interference is ruined by the Aharanov-Bohm phase
of the magnetic field. In contrast, the presence of fermionic $\mathcal{T}$
($\mathcal{T}^{2}=-1$, symplectic class), pertinent to metals with
strong spin-orbit coupling, causes disorder to be marginally irrelevant,
leads to weak anti-localization, and allows metallicity to survive
up to the thermodynamic limit at extremely weak disorder \citep{Hikami1980}.
Experimentally, weak anti-localization manifests as a peak instead
of a cusp in the magnetoresistance. Finally in 1D, disorder is relevant
and the scaling function is always negative, which physically implies
localization for infinitesimal disorder in any symmetry class.

Most of our current understanding of metallic physics is based on
the presence of at least one of inversion ($\mathcal{I}$) and time-reversal
($\mathcal{T}$) symmetries, as bulk metals that break both symmetries,
i.e., non-centrosymmetric, magnetic metals, are extremely rare. The
symmetries govern key macroscopic properties of metals via microscopic
processes such as Cooper pairing and elastic backscattering, pertinent
to superconductivity and localization, respectively. On the other
hand, largely thanks to the poor screening of electromagnetic fields,
lower dimensional systems allow phenomena that are suppressed or forbidden
in bulk materials. For instance, $\mathcal{T}$- and $\mathcal{I}$-breaking
enable a host of exotic superconducting behaviors either in systems
that are (quasi)-1D or the phenomena themselves have a directionality.
These include Majorana fermions in nanowires\,\citep{Alicea2012,Beenakker2013,Kitaev_2001,Lutchyn:2018aa,DasSarmaMajoranaSCSCJunction},
superconducting \citep{Ando:2020td,Daido2022,Lin2021,Lyu:2021wg,Narita:2022tb,Shin2021,Yuan2022,Wakatsuki2017,Wakatsuki2018,Miyasaka_2021,Zhai2022,Daido2022transition}
and Josephson diode effects\,\citep{Baumgartner:2022wr,Baumgartner_2022,Davydova2022,Diez-Merida2021,Pal:2022tm,Wu:2022wq,Halterman2022,Zhang2022Josephson,WangWangDiode},
and spontaneous supercurrents at equilibrium \citep{Samokhvalov:2021vh,Mironov2017,HosurPalaciosSupercurrent}.
This immediately raises the question, ``what are the consequences
of $\mathcal{T}$- and $\mathcal{I}$-breaking on the \emph{localization}
properties of 1D metals?''

In this work, we address this question in the simplest scenario: 1D
metals of spinless electrons with an asymmetric dispersion, $\epsilon_{k}\neq\epsilon_{-k}$,
in a disordered chemical potential. We refer to such metals as band
asymmetric metals (BAMs) and stress that they are the generic low-energy
theory of 1D metals that lack any symmetry; several examples are given
in Appendix\,\ref{sec:Model-and-realizations}. This problem technically
belongs to the unitary class; however, it differs from the usual problem
of localization in this class where disorder breaks $\mathcal{T}$
but the underlying metal does not, resulting in preserved $\mathcal{T}$
on average. In contrast, the current problem violates $\mathcal{T}$
on average too as $\mathcal{T}$ is already broken by the parent metal.
Therefore, this system is conceptually closer to a metal in a magnetic
field than to one with magnetic impurities. We study both weak and
strong localization in 1D BAMs and find that the former contains a
new physical regime while the latter enjoys a localization length
that grows parametrically with band asymmetry.

In Section \ref{sec:General-conductivity-correction}, we discuss
the WL correction to the conductivity in 1D BAMs which is followed
by the discussion of quench numerics and recursive Green's function
method calculations in Section \ref{sec:Numerics-on-zigzag-chain}
and is concluded by a discussion on possible avenues for experimental
realizations in Section \ref{sec:Experiments}. The appendices contain
the discussion of some physical models with $\mathcal{T}$- and $\mathcal{I}$-
breaking perturbations, the details of the conductivity correction
calculations, and the specifics of the recursive Green's function
(RGF) method. 

\section{General conductivity correction\label{sec:General-conductivity-correction}}

\begin{figure}
\noindent \begin{centering}
\includegraphics[scale=0.4]{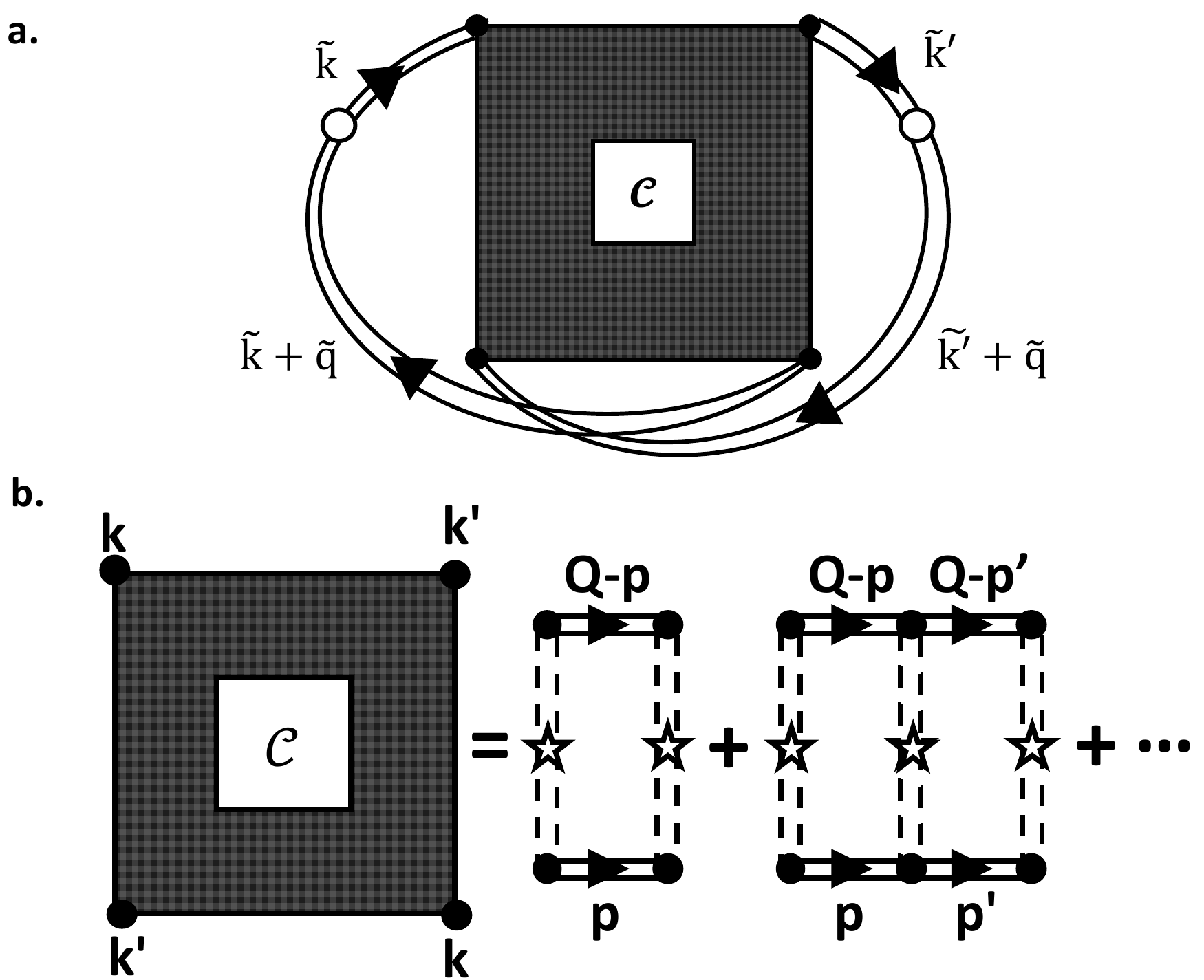}
\par\end{centering}
\begin{centering}
\includegraphics[scale=0.4]{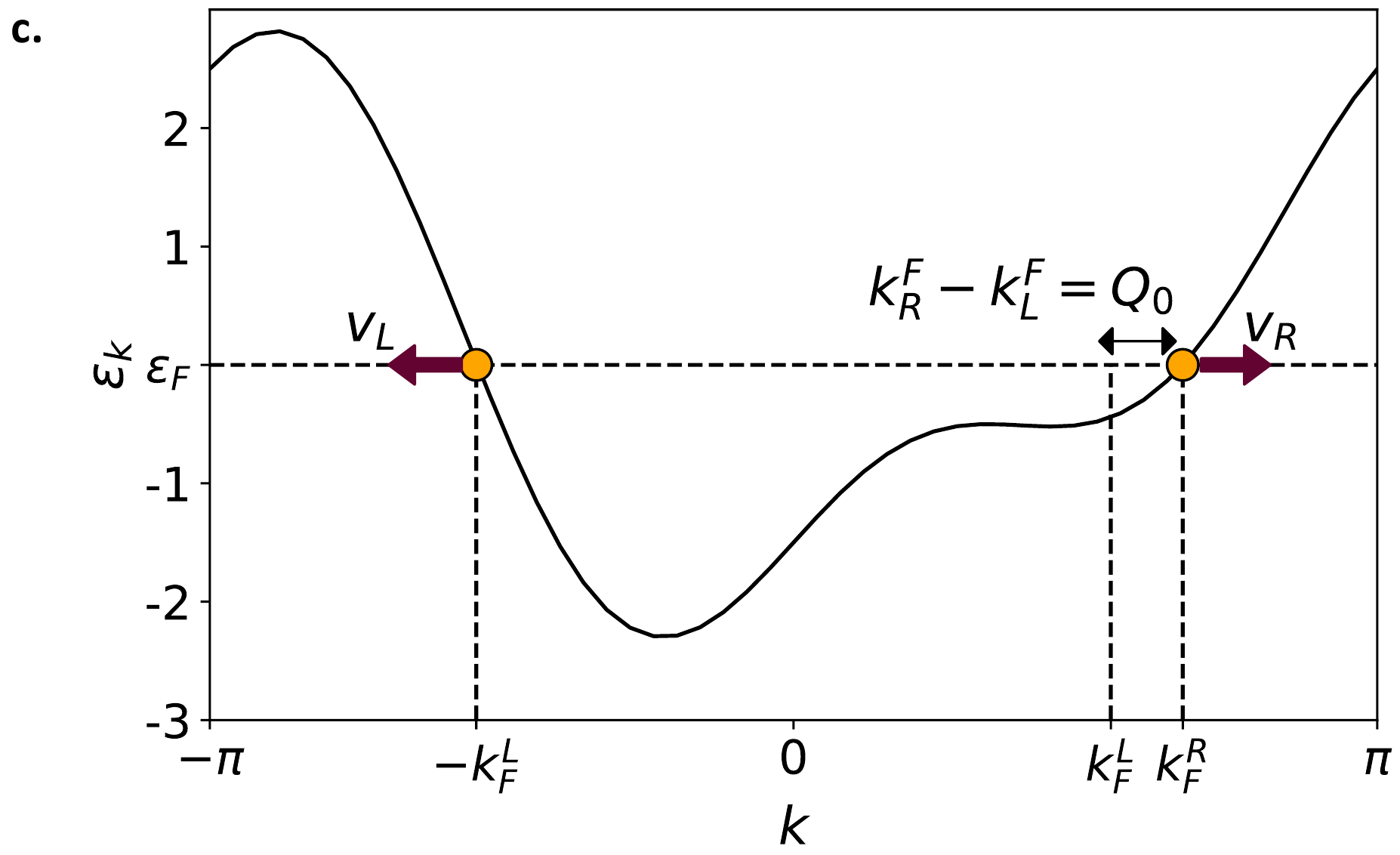}
\par\end{centering}
\caption{\label{fig:Feynman_Diagram}(a) Feynman diagrams representing the
polarization bubble for the maximally crossed (Langer-Neal) diagram,
where $\mathcal{C}$ is the resummed Cooperon propagator represented
diagrammatically by a sum of parallel impurity lines in (b). Here,
$\tilde{k}\equiv(k,i\nu_{n})$ and $\tilde{k}'\equiv(k',i\nu_{n}+i\omega_{n})$
label the top incoming and outgoing lines while $\tilde{q}\equiv(q,i\omega_{n})$
is the difference between their frequency/momentum. (c) Representative
asymmetric band showing the Fermi points and highlighting the difference
in the speeds of the left and right movers.}
\end{figure}
Our main result is a new regime of WL in 1D BAMs. Specifically, we
show that the WL correction to the conductivity in 1D BAMs is given
by:

\begin{align}
\sigma^{\text{WL}} & \approx-\frac{e^{2}}{\hbar}\frac{2\pi v\tau_{0}}{\sqrt{\frac{l}{l_{\phi}}+\frac{\delta v^{2}}{4v^{2}}}}.\label{eq:sigma-WL}
\end{align}
for $l/l_{\phi},|\delta v|/v\ll1$, where $v=(v_{L}+v_{R})/2$ is
the average speed of the left and right movers, $\delta v=v_{R}-v_{L}$
is the difference in speeds, $l$ is the mean free path, $l_{\phi}$
is a phenomenological phase coherence length that is typically governed
by inelastic scattering and thermal decoherence, and $\tau_{0}$ is
the quasiparticle lifetime calculated in the Born approximation. We
have assumed a single pair of counterpropagating modes for simplicity.
Eq.\,(\ref{eq:sigma-WL}) shows two distinct regimes: for $\sqrt{\frac{l}{l_{\phi}}}\ll\frac{|\delta v|}{2v}$
and $\sqrt{\frac{l}{l_{\phi}}}\gg\frac{|\delta v|}{2v}$, dephasing
is dominated by band asymmetry and inelastic scattering, respectively.
Thus, symmetric metals with $\delta v=0$ fall in the latter regime
and have $\sigma^{\text{WL}}\propto-\sqrt{l_{\phi}}$ \citep{bruus_flensberg_2020},
which diverges at zero temperature in the absence of inelastic scattering
processes. Intuitively, right and left moving waves at a given speed
have equal and opposite momenta. Therefore, they form a perfect standing
wave and enhance localization. In contrast, if band asymmetry is large
enough {[}Fig.\,\ref{fig:Feynman_Diagram}\,(c){]}, the standing
wave heuristically melts into an interference pattern with net drift.
Eq.\,(\ref{eq:sigma-WL}) predicts this for $\sqrt{\delta v/v}\gg l/l_{\phi}$;
then $\sigma^{\text{WL}}$ remains finite as $l_{\phi}$ diverges
and depends on disorder only through $\tau_{0}$. 

To arrive at Eq.\,(\ref{eq:sigma-WL}), we begin by assuming random
chemical potential quenched disorder and considering the effect of
band asymmetry on $\tau_{0}^{-1}=2\text{Im}\Sigma(i0^{+})$ in the
Born approximation, where $\Sigma(z)$ is the complex frequency dependent
self-energy. Although the BAM has unequal Fermi momenta for left and
right movers, $\left|k_{F}^{L}\right|\neq\left|k_{F}^{R}\right|$
{[}Fig.\,\ref{fig:Feynman_Diagram}\,(c){]}, $\tau_{0}$ depends
on the band structure only through the density of states at the Fermi
level. As a result, we find that it changes quantitatively, but not
qualitatively, as the bands turn asymmetric. Physically, this means
band asymmetry does not qualitatively affect classical transport,
i.e., transport in the regime where quantum interference effects vanish
and probabilities rather than amplitudes for different Feynman paths
add. Thus, one must transcend the Born approximation and consider
appropriate vertex corrections to see the qualitative effects of band
asymmetry.

In a $\mathcal{T}$-symmetric system, the vertex corrections that
survive disorder-averaging consist of maximally crossed diagrams,
illustrated in Fig.\,\ref{fig:Feynman_Diagram}\,(a,b). Thus, we
evaluate the polarization bubble with these corrections following
standard procedure \citep{coleman_2019,bruus_flensberg_2020} to obtain
$\sigma^{\text{WL}}$; see Appendix\,\ref{sec:Detailed-calculation}
for full details of the calculation. When computed for metals under
a small orbital magnetic field, these corrections yield the well-known
experimental signatures of WL in magnetoresistance. In the present
context, fortunately, the calculation is simpler since band asymmetry
is a non-singular perturbation unlike an orbital magnetic field. In
particular, all momentum integrals here can be elegantly done by contour
methods once we note that the integrals are dominated by regions near
the Fermi points and linearize the dispersion around these points.
Moreover, the dominant contributions to WL are captured by the retarded-advanced
Cooperon propagator since the phenomenon effectively arises from interference
between forward and backward time evolution of the electron wavefunction.
Under the linear approximation, we find the retarded-retarded Cooperon
propagator to exactly vanish. Linearization also naturally introduces
the Fermi velocities $v_{L,R}$ into the calculation and lets us package
the band asymmetry into a single dimensionless parameter, $\delta v/2v=(v_{R}-v_{L})/(v_{L}+v_{R})$.
Finally, we resum the Dyson series for the Cooperon propagator and
calculate the polarization bubble for conductivity to obtain $\sigma^{\text{WL}}$.
Along the way, we include Markovian inelastic scattering into the
calculation via a phenomenological phase decoherence probability $e^{-l/l_{\phi}}$
between elastic scattering events. This yields the result, Eq.\,(\ref{eq:sigma-WL}),
for $l\ll l_{\phi}$ and $\delta v\ll v$.

The dependence of $\sigma^{\text{WL}}$ on band asymmetry only through
$\delta v$ indicates that the above phenomena appear in a wide range
of physical systems. In Appendix\,\ref{sec:Model-and-realizations},
we describe several systems with $\mathcal{T}$- and $\mathcal{I}$-
breaking perturbations, some of which are dynamically tunable and
have seen experimental realizations, where we expect a suppression
of WL. In the next section, we focus on a lattice model and study
localization in it numerically.

\section{Numerics on zigzag chain\label{sec:Numerics-on-zigzag-chain}}

\begin{figure}
\begin{centering}
\includegraphics[scale=0.45]{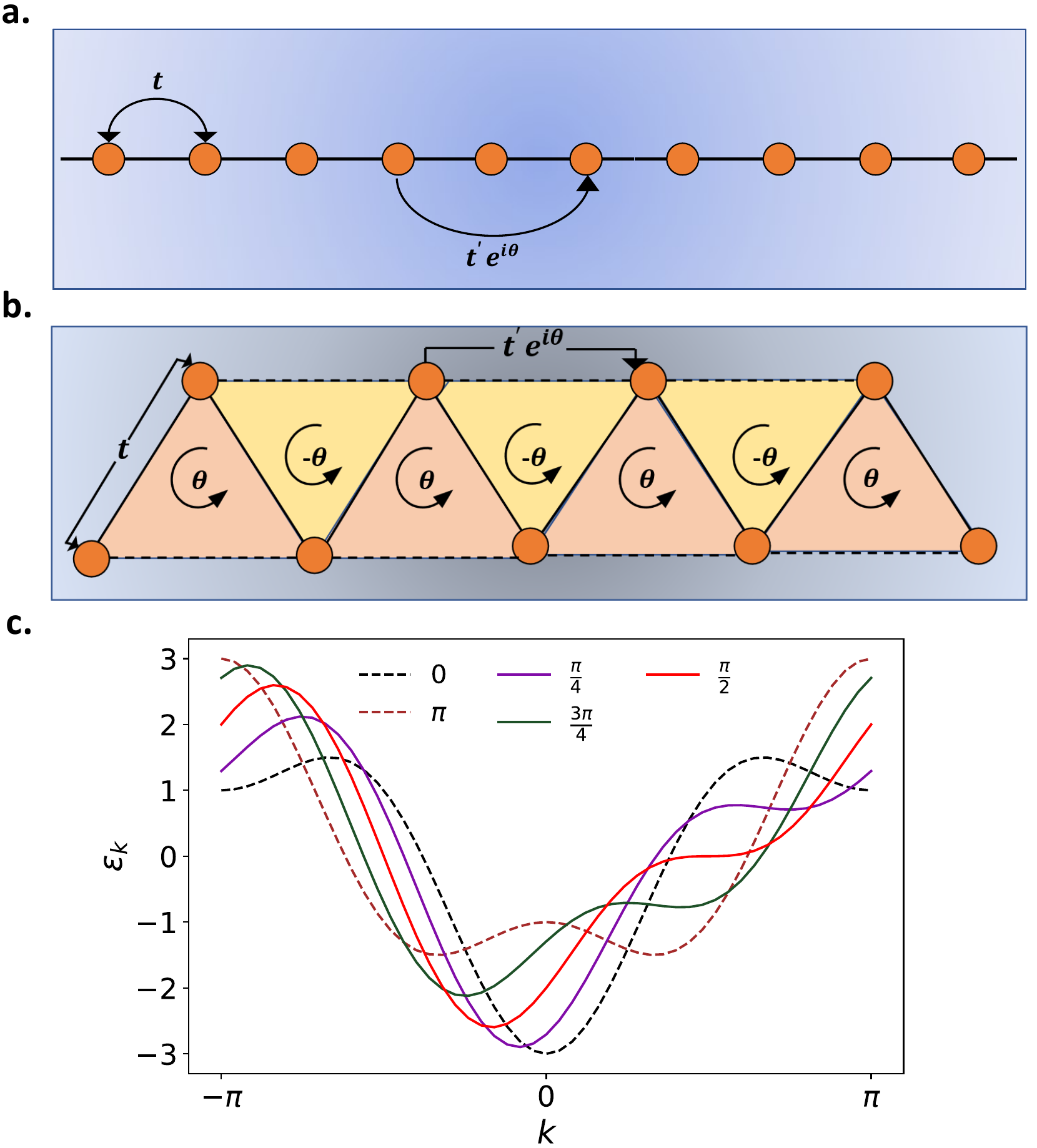}
\par\end{centering}
\caption{\label{fig:ToyModel_Dispersion}(a) A schematic representation of
the 1D tight-binding model with real NN and complex NNN hopping. (b)
Zigzag chain representation of the same model with alternating flux
passing through adjacent triangles. (c) Graphical representation of
the dispersion relation $\epsilon_{k}$. Symmetric cases ($\theta=0$
and $\theta=\pi$) and the asymmetric cases are represented by dashed
and solid lines respectively.}
\end{figure}
\begin{figure*}
\begin{centering}
\includegraphics[scale=0.4]{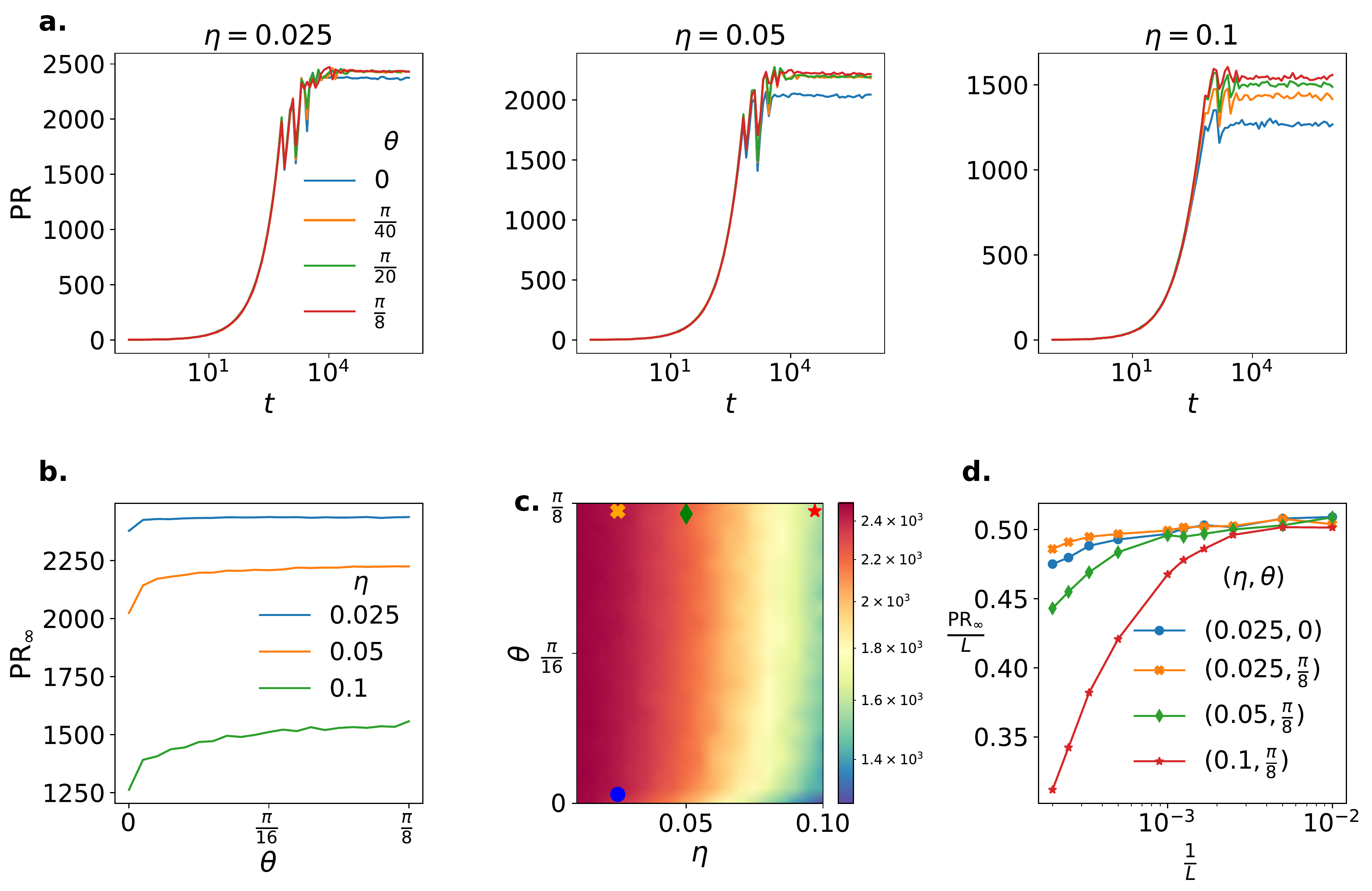}
\par\end{centering}
\caption{\label{fig:Quench}PR saturation with time for a system with $L=5000.$
(a) Time evolution of PR up to $t=10^{6}$ for various $\theta$,
for a given $\eta$. (b) The long-time saturation value of PR, PR$_{\infty}$,
as a function of $\theta$ for various $\eta$. (c) Color plot showing
the variation of $\mathrm{PR}_{\infty}$ with $\theta$ and $\eta$.
(d) Finite-size scaling behavior of $\frac{\mathrm{PR}_{\infty}}{L}$
vs $\frac{1}{L}$ up to $L=5000$ for some $(\eta,\theta)$ pairs
marked in (c).}
\end{figure*}

To substantiate the analytics, we study localization numerically on
a tight-binding lattice model of spinless fermions {[}Fig.\,\ref{fig:ToyModel_Dispersion}{]}
described by

\begin{align}
H & =-t\sum_{i}c_{i}^{\dagger}c_{i+1}-t'e^{i\theta}\sum_{i}c_{i}^{\dagger}c_{i+2}+h.c.\nonumber \\
 & \qquad+\sum(\varepsilon_{i}-\mu)c_{i}^{\dagger}c_{i},
\end{align}
where $c_{i}$ and $c_{i}^{\dagger}$ are fermionic annihilation and
creation operators at the lattice site `$i$'. Also, $t$, $t'$,
$\theta$, $\mu$, and $\varepsilon_{i}$ represent the nearest neighbor
(NN) hopping strength, next-nearest neighbor (NNN) hopping strength,
$\mathcal{T}$- and $\mathcal{I}$-breaking hopping phase, chemical
potential, and the on-site disorder potential respectively. The dispersion
in the absence of disorder is

\begin{equation}
\epsilon_{k}=-2[t\cos(k)+t'\cos(2k+\theta)]-\mu,\label{eq:Dispersion}
\end{equation}
which shows band asymmetry for generic values of $\theta\neq0,\pi$
{[}Fig.\,\ref{fig:ToyModel_Dispersion}\,(c){]}.

This model can also be viewed as a zigzag chain with triangular plaquettes
{[}Fig.\,\ref{fig:ToyModel_Dispersion}\,(b){]}. The NN hopping
terms form the two sides of the triangles and NNN ones are across
the bases. These triangular plaquettes have a total phase of $\pm\theta$
associated with them corresponding to the total phase picked up by
a particle while hopping anti-clockwise along the edges. As we can
see in Fig.\,\ref{fig:ToyModel_Dispersion}\,(b), adjacent triangles
have opposite fluxes passing through them. However, we note that despite
having a zigzag chain representation, the system has a 1-site unit
cell and $H$ is invariant under unit translation, $i\to i+1$. Throughout
this paper, we consider $t=-1$, $t'=-0.5$, $\mu=-1$, and draw $\varepsilon_{i}$
from a uniform distribution $[-\eta,\eta]$.

\begin{figure*}
\begin{centering}
\includegraphics[scale=0.4]{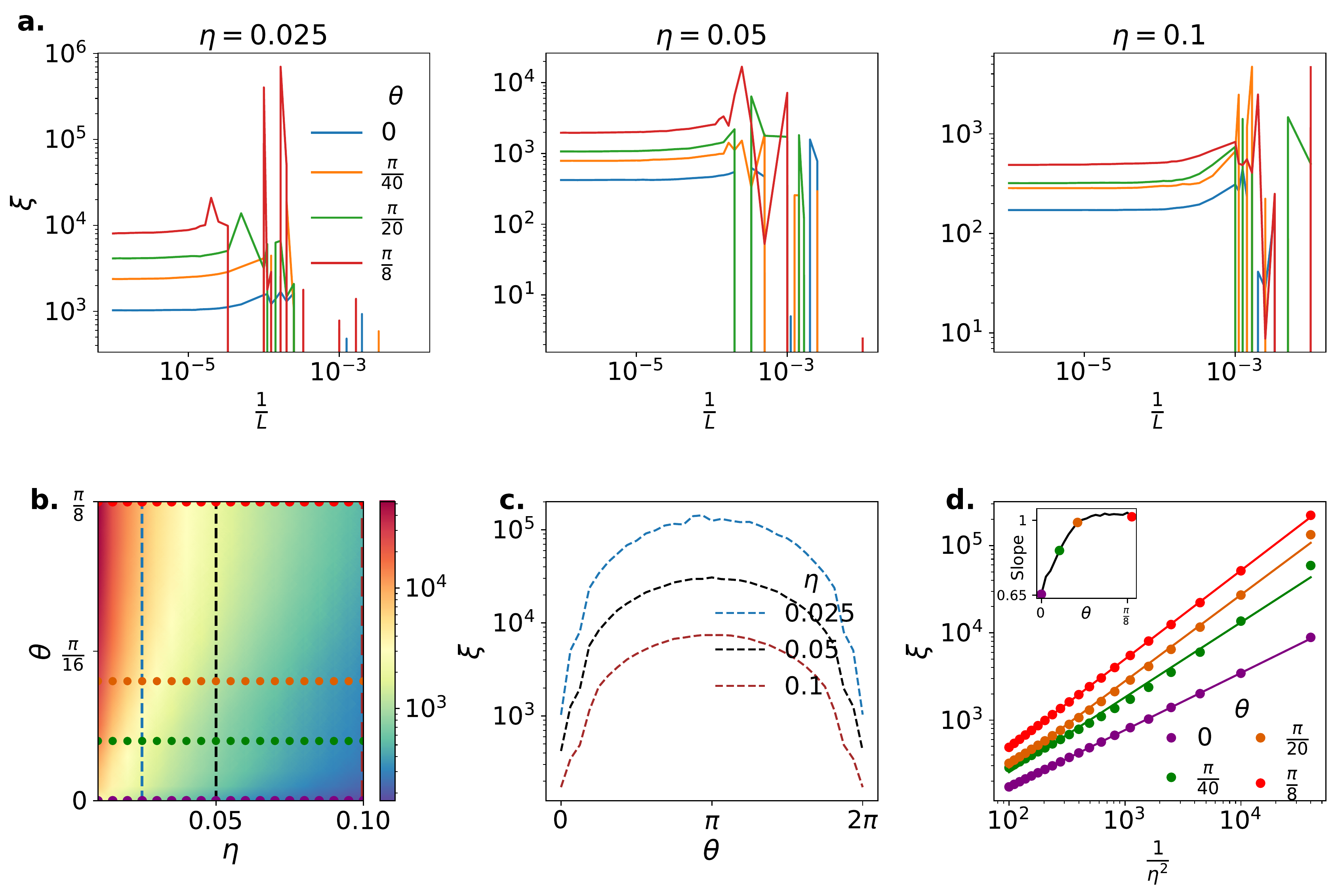}
\par\end{centering}
\caption{\label{fig:RGF}Localization length $(\xi)$ calculations using recursive
Green's function method up to $L=10^{6}$. (a) Variation of $\xi$
with $\theta$ for a given $\eta$ as a function of $\frac{1}{L}$.
(b) Color plot showing the variation of $\xi$ with $\theta$ and
$\eta$. (c) $\xi$ as a function of $\theta$ for various $\eta$.
(d) $\xi$ as a function of $\frac{1}{\eta^{2}}$ for various $\theta$.
The dots represent the simulation data and the solid lines represent
the power law fit. Inset shows the slope approaching $1$ as $\theta$
increases.}
\end{figure*}

As a first diagnostic tool, we calculate the participation ratio (PR)\textbf{
}defined as 
\begin{equation}
\text{PR}=\frac{1}{\sum_{i}|\psi_{i}|^{4}},\label{eq:PR}
\end{equation}
for a tight-binding wavefunction $\psi$. It is a measure of the number
of states a particle is distributed over. For a particle localized
on only one site, $\text{PR}=1$, while a particle evenly distributed
over $L$ sites has $\text{PR}=L$. In general, a finite (vanishing)
value of PR$/L$ as $L\to\infty$ indicates delocalization (localization).

To study localization in our system, we numerically calculate the
spread of wavefunctions starting from one that is localized on two
neighboring sites, $\frac{|i\rangle+|i+1\rangle}{\sqrt{2}}$, $|i\rangle$
being the state corresponding to the particle localized on site `$i$'.
We have chosen this particular initial condition because its energy
lies at the Fermi level for $\mu=-1$ and the effect of the suppression
of weak localization is seen prominently at finite $\mu$. We choose
the sites in the middle of the chain but the location does not matter
as we are using periodic boundary conditions in all our quantum quench
calculations. We evolve the system up to $t=10^{6}$ and perform a
disorder average of 100 disorder realizations in most cases. 

In Fig.\,\ref{fig:Quench}\,(a), we present the time evolution of
PR for different $\eta$ and $\theta$ in a system with $L=5000$.
The wavefunction of the particle starting in the middle of the lattice
spreads over the system initially, resulting in an increase in PR
with time. However, the spread does not continue indefinitely and
PR saturates after some time. Increasing $\eta$ reduces the saturation
value of PR, which is consistent with WL; however, a non-zero $\theta$
suppresses this effect resulting in a higher PR saturation value compared
to the symmetric case {[}Fig\,\ref{fig:Quench}\,(c){]}. We note
that the suppression of WL increases monotonically and rapidly as
$\theta$ increases away from 0 {[}Fig.\,\ref{fig:Quench}\,(b){]}.
We then perform finite size scaling of PR saturation values for various
$(\theta,\eta)$ pairs. For small $L$, the PR increases linearly
with $L$. As we argue below, this is because the localization length
$\xi>L$ at these system sizes. However, $\frac{\mathrm{PR}_{\infty}}{L}$
decreases as we go towards larger $L$ {[}Fig.\,\ref{fig:Quench}\,(d){]},
suggesting that 1D free-fermions systems might localize in the thermodynamic
limit at arbitrarily small disorders even in the presence of band
asymmetry.

The last statement is difficult to check using exact diagonalization
for very weak disorder as $\xi$ can be much larger than the system
sizes this method can access. Fortunately, iterative algorithms such
as transfer matrix \citep{Numerical_Anderson} and recursive Green's
function (RGF) method \citep{RGF,RGF_Localization_Length,RGF_Original_Kramer_MacKinnon}
exist which can access significantly larger system sizes. Thus, we
complement our quench numerics by explicitly computing $\xi$ using
the RGF method. We explore various $\eta$ and $\theta$ and system
sizes up to $L=10^{6}$, and average over 100 realizations. The details
of RGF method for our system are presented in Appendix\,\ref{sec:Recursive-Green's-function}.

We choose the Fermi energy for $\mu=-1$ as before to see the suppression
of WL prominently. In Fig.\,\ref{fig:RGF}\,(a), we present the
iterations of the RGF method calculations as a function of system
size. After some fluctuations at small sizes, $\xi$ clearly saturates
to a constant value that defines the localization length in the thermodynamic
limit for that particular $(\theta,\eta)$ pair. Fig.\,\ref{fig:RGF}\,(b)
presents $\xi$ as a function of both $\eta$ and $\theta$ while
Fig.\,\ref{fig:RGF}(c,d) show cuts through this plot. For a particular
$\eta$, $\xi$ increases monotonically with increasing $\theta$
{[}Fig.\,\ref{fig:RGF}\,(c){]} and decreases with increasing $\eta$
{[}Fig.\,\ref{fig:RGF}\,(d){]}. We see in Fig.\,\ref{fig:RGF}\,(c)
that asymmetry increases $\xi$ but shows signs of saturation rather
than divergence, suggesting that the system is still localized. 

In the regime of weak $\eta$ and $\theta=0$, perturbation theory
\citep{Localization_Theory_Experiment,THOULESS1984} predicts $\xi\propto\frac{1}{\eta^{2}}$.
By employing similar arguments using Green function $G(E)=\intop_{k}(E-\epsilon_{k})^{-1}$
for $\theta\neq0$, we get 
\begin{align}
\frac{1}{\xi} & =-\frac{\eta^{2}}{12}\intop^{E}G\frac{dG}{dE},
\end{align}
 where 

\begin{align}
\intop^{E}G\frac{dG}{dE} & =-\intop^{E}\intop_{k,k'}\frac{1}{E-\epsilon_{k}}\frac{1}{(E-\epsilon_{k'})^{2}}\\
 & =-\intop_{k,k'}A_{k,k'}\ln(E-\epsilon_{k})+\frac{B_{k,k'}}{E-\epsilon_{k'}},
\end{align}
Provided the above integral is finite or properly regularized, the
scaling behaviour $\xi\propto\frac{1}{\eta^{2}}$ holds true even
for finite values of $\theta$. We have verified this relationship
between $\xi$ and $\frac{1}{\eta^{2}}$ by performing a power law
fit (solid lines) on the simulation data points (dots) {[}Fig.\,\ref{fig:RGF}\,(d){]},
where we observe the slope approaching $1$ as $\theta$ increases
{[}Fig.\,\ref{fig:RGF}\,(d) (inset){]}. Since the scaling $\xi\propto\frac{1}{\eta^{2}}$
is well-known at $\theta=0$ in the thermodynamic limit \citep{Localization_Theory_Experiment,THOULESS1984},
the deviation from this scaling is presumably due to finite size effects.
This suggests that finite size effects, surprisingly, are smaller
for $\theta\neq0$.

PR and $\xi$ both measure the degree of localization in a system,
with PR quantifying the number of sites the wavefunction is spread
over and $\xi$ being the characteristic length scale over which the
wave function decays. Therefore, we observe consistent trends in PR
and $\xi$ with varying $\eta$ and $\theta$. Notably, the RGF method
calculations yield large values of $\xi$ for certain parameters,
surpassing the system size used in exact diagonalization. It is important
to note here that these calculations correspond to two different regimes.
The analytical result {[}Eq.\,(\ref{eq:sigma-WL}){]} is valid, and
the PR calculations {[}Fig.\,\ref{fig:Quench}{]} are done in the
regime $\xi>L$ where a finite system is delocalized, whereas the
RGF method calculations {[}Fig.\,\ref{fig:RGF}{]} are performed
in the regime $L>\xi$ which is smoothly connected to the thermodynamic
limit. The trends in these numerical calculations align with the analytical
result {[}Eq.\,(\ref{eq:sigma-WL}){]} that the conductivity increases
due to asymmetry, though a finite $\xi$ {[}Fig.\,\ref{fig:RGF}{]}
implies that the conductivity is $0$ by definition in the thermodynamic
limit. 

\section{Experiments\label{sec:Experiments}}

The suppression of localization due to band asymmetry can be probed
in 1D metallic wires with Rashba spin-orbit coupling using a magnetic
field. As described in Appendix\,\ref{sec:Model-and-realizations},
the dispersion is symmetric in the absence of a magnetic field due
to $\mathcal{T}$. A weak magnetic field will break $\mathcal{T}$,
turn the bands asymmetric, and should enhance the conductivity. We
emphasize that the enhancement in 1D is due to the Zeeman effect of
the magnetic field, and is distinct from the usual suppression of
WL due to the Aharanov-Bohm effect of an orbital field in 2D, 3D and
even in 1D wires with a finite width \citep{Liang:2009aa,Beenakker1988}.

Concrete realizations of our model {[}Fig.\,\ref{fig:ToyModel_Dispersion}{]}
may also be achieved in synthetic \citep{Zigzag_Experiment} and optical
\citep{Alternate_Flux} lattices, which offer high tunability of hopping
amplitudes, disorder and fluxes using artificial gauge fields. Moreover,
unlike solids, these platforms naturally lack phonons and uncontrolled
disorder. With increasing flux as illustrated in Fig.\,\ref{fig:ToyModel_Dispersion},
we predict a greater spread of an initially local wavefunction. In
particular, the localization length extracted from the long-time density
profiles should increase as the flux increases. While the Aharanov-Bohm
flux plays a key role in this realization, its main role is to break
the $k\to-k$ symmetry of the bands. Indeed, the average flux is zero,
which distinguishes it from usual studies of WL in uniform magnetic
fields in solid state systems.

\section{Conclusions}

We have shown that in 1D metals where time reversal and inversion
symmetry are broken, dubbed band asymmetric metals, weak localization
is suppressed due to the asymmetry in velocities of left and right
movers. Heuristically, the formation of perfect standing waves due
to quantum interference between time-reversed paths, leading to weak
localization, is disrupted due to this asymmetry. The analytical results
are validated by the numerical calculations of the participation ratio
and localization length such that there is an increase in conductivity,
participation ratio, and localization length with increasing band
asymmetry, indicating a tendency to delocalize. Metallic nanowires
with strong spin-orbit coupling and tunable synthetic and optical
lattices with controlled disorders may be convenient platforms for
experimentally investigating the impact of band asymmetry on the localization
properties of disordered systems.
\begin{acknowledgments}
P.H. was supported by the Department of Energy under grant no. DE-SC0022264.
P.H. would like to thank Pouyan Ghaemi, Joseph Maciejko, and Hridis
Pal for useful discussions. We acknowledge the National Supercomputing
Mission (NSM) for providing computing resources of \textquoteleft PARAM
Shivay\textquoteright{} at the Indian Institute of Technology (BHU),
Varanasi, which is implemented by C-DAC and supported by the Ministry
of Electronics and Information Technology (MeitY) and Department of
Science and Technology (DST), Government of India. Some of the calculations
have been performed using the Quspin library \citep{Quspin-1,Quspin-2}.
\end{acknowledgments}

\begin{widetext}

\appendix

\section{Models and realizations\label{sec:Model-and-realizations}}

Here, we discuss three physical models with $\mathcal{T}$- and $\mathcal{I}$-
breaking perturbations and calculate the asymmetry in speeds of left
and right movers to leading order.

\subsection{Zigzag chain}

Considering the zigzag chain dispersion relation {[}Eq.\,(\ref{eq:Dispersion}){]},
for the symmetric case $(\theta=0,\pi)$ we find equal (in magnitude)
and opposite (in direction) Fermi momenta
\begin{equation}
\pm k_{F}=\pm\begin{cases}
\cos^{-1}\left(\frac{-2t+\sqrt{4t^{2}-16t'(\mu-2t')}}{8t'}\right) & \theta=0\\
\cos^{-1}\left(\frac{2t-\sqrt{4t^{2}+16t'(\mu+2t')}}{8t'}\right) & \theta=\pi
\end{cases},
\end{equation}
 and equal Fermi speeds $v_{R}=-v_{L}\equiv v_{F}$ where
\begin{equation}
v_{F}=\left|2t\sin k_{F}+4t'\sin2k_{F}\right|.
\end{equation}
The perturbative correction due to small asymmetry $\theta$ to the
leading order breaks $\mathcal{T}$and $\mathcal{I}$ symmetries to
give $k_{F}^{R/L}=k_{F}\pm\delta$, where 
\begin{equation}
\delta=\frac{-2t'\cos k_{F}}{t+4t'\cos k_{F}}\theta,
\end{equation}
and 
\begin{align}
v_{R/L} & =v_{F}\mp\frac{4tt'\sin^{2}k_{F}}{t+4t'\cos k_{F}}\theta,
\end{align}
showing the difference in the magnitude of Fermi speeds, $\delta v\propto\theta$.

\subsection{Cubic perturbation}

For a continuum model with a cubic perturbation, the dispersion relation
is given by

\begin{align}
\epsilon_{cubic} & =\frac{k^{2}}{2m}+\alpha k^{3}+\beta k^{4}-\mu,
\end{align}
where $\alpha k^{3}$ is the $\mathcal{T}$- and $\mathcal{I}$- breaking
perturbation and $\beta k^{4}$ with $\beta>0$ keeps $\epsilon_{cubic}$
positive in the limit $k\to\pm\infty$. We treat the quadratic and
quartic terms in the dispersion as the unperturbed system and calculate
the correction due to the cubic term.\textbf{ }Similar to the zigzag
chain, the symmetric case $(\alpha=0)$ has equal and opposite Fermi
momenta 
\begin{equation}
\pm k_{F}=\pm\sqrt{\frac{-1+\sqrt{1+16\mu\beta m^{2}}}{4m\beta}}
\end{equation}
and equal Fermi speeds 
\begin{equation}
v_{F}=\frac{k_{F}}{m}+4\beta k_{F}^{3}
\end{equation}
The cubic perturbation gives $k_{F}^{R/L}=k_{F}\pm\delta$, where
$\delta=-\alpha k_{F}^{2}m$ to leading order, and 
\begin{equation}
v_{R/L}\approx v_{F}\pm2k_{F}^{2}\left[1-\frac{6\beta k_{F}^{2}}{m}\right]\alpha,
\end{equation}
resulting in the difference in the magnitude of Fermi speeds, $\delta v\propto\alpha$. 

\subsection{Spin-orbit coupling and magnetization}

Here, we consider a wire along $x$ with a Zeeman field $B_{y}$ and
Rashba spin-orbit coupling $\lambda$. Its Hamiltonian is
\begin{equation}
H=\frac{k^{2}}{2m}-\mu+\lambda k\sigma_{y}-\gamma\mu_{B}B_{y}\frac{\sigma_{y}}{2},
\end{equation}
where $\gamma$ and $\mu_{B}$ are the gyromagnetic ratio and Bohr
magneton, respectively, and $\sigma_{y}=\pm1$ refers to spin along
$y$. For each value of $\sigma_{y}$, there exists a right mover
and a left mover at the Fermi level, resulting in a total of four
Fermi points. For this system, there are two ways to obtain a symmetric
band structure. The first is by switching off the magnetic field $(B_{y}=0)$,
which yields two pairs of equal and opposite Fermi momenta:
\begin{equation}
\left|k_{F}\right|=m\left|-\sigma_{y}\lambda\pm\sqrt{\lambda^{2}+\frac{2\mu}{m}}\right|,
\end{equation}
and equal Fermi speeds 
\begin{equation}
v_{F}=\sqrt{\lambda^{2}+\frac{2\mu}{m}}.
\end{equation}
In this case, band symmetry exists between a right-mover with spin
$\sigma_{y}$ and left-mover with spin $-\sigma_{y}$ due to $\mathcal{T}$.
The second way to obtain a symmetric dispersion is by suppressing
spin-orbit coupling $(\lambda=0)$. This, too, gives a pair of equal
and opposite Fermi momenta for each value of $\sigma_{y}$ due to
$\mathcal{I}$: 
\begin{equation}
K_{F,\sigma_{y}}=\sqrt{m(2\mu+\gamma\mu_{B}B_{y}\sigma_{y})},
\end{equation}
and equal Fermi speeds
\begin{equation}
V_{F,\sigma_{y}}=\sqrt{\frac{2\mu+\gamma\mu_{B}B_{y}\sigma_{y}}{m},}
\end{equation}
When both $B_{y}$ and $\lambda$ are non-zero, neither the Fermi
points nor the velocities appear in equal and opposite pairs. Their
values now are, 
\begin{equation}
k_{\sigma_{y}}^{R/L}=m\left(-\sigma_{y}\lambda+\sqrt{\lambda^{2}+\frac{2\mu+\sigma_{y}\gamma\mu_{B}B_{y}}{m}}\right),
\end{equation}
\begin{equation}
v_{\sigma_{y}}^{R/L}=\sqrt{\lambda^{2}+\frac{2\mu+\sigma_{y}\gamma\mu_{B}B_{y}}{m}},
\end{equation}

\section{Detailed calculation of $\sigma^{\text{WL}}$\label{sec:Detailed-calculation}}

\subsection{Self-energy}

We consider a disorder potential $U(x)=\sum_{n}\mathcal{U}(x-R_{n})$
with $\bar{U(x)}=0$ and $\bar{U(x)U(x')}=n_{i}u_{0}^{2}\delta(x-x')$
where the bar denotes disorder-average and $n_{i}$ is the impurity
density, and begin the analysis by considering the effect of band
asymmetry on the self-energy as a function of complex frequency $\Sigma(z)$
\citep{bruus_flensberg_2020,coleman_2019}:
\begin{equation}
\Sigma(z)=n_{i}u_{0}^{2}\intop_{-\pi}^{\pi}\frac{dk}{2\pi}\frac{1}{z-\epsilon_{k}},
\end{equation}
Analytically continuing $z$ within a half-plane, $z\to\varepsilon+i0^{+}\text{sgn}[\text{Im}(z)]$
and absorbing $\text{Re}\Sigma$ into a redefinition of $\mu$, we
get a lifetime from the imaginary part:
\begin{align}
\frac{1}{\tau(\varepsilon)} & =2\pi n_{i}u_{0}^{2}\intop_{k}\delta(\varepsilon-\epsilon_{k})
\end{align}
To simplify the analysis, let us assume there is a single left-mover
at each $\varepsilon$ with speed $v_{L}(\varepsilon)$ and a single
right-mover with speed $v_{R}(\varepsilon)$. Then,
\begin{align}
\frac{1}{\tau(\varepsilon)} & =n_{i}u_{0}^{2}\left(\frac{1}{v_{L}(\varepsilon)}+\frac{1}{v_{R}(\varepsilon)}\right),
\end{align}
where substituting the density of states per unit length $g(\varepsilon)=\frac{1}{2\pi}\left(\frac{1}{v_{L}(\varepsilon)}+\frac{1}{v_{R}(\varepsilon)}\right)$,
we get: 
\begin{equation}
\frac{1}{\tau_{0}(\varepsilon)}=2\pi n_{i}u_{0}^{2}g(\varepsilon)
\end{equation}
Clearly, the Born lifetime depends only on the mean inverse speed
and is not affected qualitatively by the velocity asymmetry. Nevertheless,
it is convenient to separate the speeds into their average and differences,
$v_{L}=v-\delta v/2$, $v_{R}=v+\delta v/2$. This gives
\begin{equation}
\frac{1}{\tau_{0}(\varepsilon)}=\frac{2n_{i}u_{0}^{2}v(\varepsilon)}{v^{2}(\varepsilon)-\delta v^{2}(\varepsilon)/4}\label{eq:tau0}
\end{equation}
Here, $G^{R}(k,\varepsilon)$, $G^{A}(k,\varepsilon)$, and $\mathcal{G}(k,i\nu_{n})$
are the usual retarded, advanced, and Matsubara electron Green's functions.

\subsection{Cooperon propagator}

To determine the weak localization correction to the conductivity,
$\sigma^{\text{WL}}(\tilde{q})$, we need to calculate the polarization
bubble due to the maximally crossed or Langer-Neal diagrams that capture
constructive interference between time-reversed paths \citep{bruus_flensberg_2020,coleman_2019}.
The bubble is given by
\begin{equation}
\Pi^{WL}(\tilde{q})=-\intop_{\tilde{k},\tilde{k}^{\prime}}v_{k+q/2}v_{k'+q/2}\mathcal{G}(\tilde{k})\mathcal{G}(\tilde{k}+\tilde{q})\mathcal{C}(\tilde{k},\tilde{k}^{\prime},\tilde{q})\mathcal{G}(\tilde{k}^{\prime})\mathcal{G}(\tilde{k}^{\prime}+\tilde{q}),
\end{equation}
where $\tilde{k}\equiv(k,i\nu_{n})$, $\tilde{q}\equiv(q,i\omega_{n})$,
$\intop_{\tilde{k}}\equiv T\sum_{ik_{n}}\int\frac{dk}{2\pi}$, and
$\mathcal{C}$ is the resummed Cooperon propagator that is represented
diagrammatically by a sum of parallel impurity lines {[}Fig.\,\ref{fig:Feynman_Diagram}\,(a,b){]}.
We use the notation $\mathcal{C}(\tilde{k},\tilde{k}^{\prime},\tilde{q})$,
where $\tilde{k}$ and $\tilde{k}^{\prime}$ label the top incoming
and outgoing lines while $\tilde{q}$ is the difference between the
frequency/momentum of the bottom outgoing line and the top incoming
line.

Some simplifications occur or can be justifiably made while doing
these calculations. Since we will eventually be interested in the
dc limit, so we can set $q=0$. Also, impurity lines after disorder-averaging
behave like interactions that conserve frequency, so the top and bottom
fermion lines have frequencies $i\nu_{n}$ and $i\nu_{n}+i\omega_{n}$
for every diagram in the Dyson series for $\mathcal{C}$, or in short,
$\mathcal{C}\propto\frac{1}{T}\delta_{\nu_{n},\nu_{n}^{\prime}}$.
Finally, for short-range impurities, the scattering potential is momentum-independent.
This ensures that $C(k,k',q=0)$ only depends on $k+k'\equiv Q$.

The Dyson series for $\mathcal{C}$ can now be resummed, and yields:
\begin{equation}
\mathcal{C}(Q;i\nu_{n}+i\omega_{n},i\nu_{n})=\frac{n_{i}^{2}u_{0}^{4}\int_{p}\mathcal{G}(Q-p,i\nu_{n}+i\omega_{n})\mathcal{G}(p,i\nu_{n})}{1-n_{i}u_{0}^{2}\int_{p}\mathcal{G}(Q-p,i\nu_{n}+i\omega_{n})\mathcal{G}(p,i\nu_{n})},
\end{equation}
where we have written the frequency and momentum arguments separately
in the Green's functions. 

\subsection{Contour integrals}

This calculation can be carried out in two steps. 

\subsubsection{Frequency integral}

Since two complex frequencies are involved, there are two branch cuts,
at $\text{Im}z=0$, and $\text{Im}z=-i\omega_{n}$ where $i\nu_{n}\to z$.
We first perform the frequency integration by summing over one of
the two frequencies $(i\nu_{n})$, doing a Taylor expansion around
the other $(\omega),$ and calculating the correction to the conductivity
\citep{bruus_flensberg_2020,coleman_2019}, $\sigma^{\text{WL}}=-\lim_{\omega\to0}\frac{1}{\omega}\text{Im}\Pi^{\text{WL}}(\omega)$.
This gives
\begin{equation}
\sigma^{\text{WL}}=-\intop_{k,k^{\prime},\varepsilon}v_{k}v_{k'}f'(\varepsilon)\left[G^{A}(k,\varepsilon)G^{A}(k',\varepsilon)C^{RA}(Q;\varepsilon,\varepsilon)-G^{R}(k,\varepsilon)G^{R}(k',\varepsilon)C^{RR}(Q;\varepsilon,\varepsilon)\right]G^{R}(k',\varepsilon)G^{R}(k,\varepsilon)
\end{equation}
where 
\begin{align}
C^{RA}(Q;\varepsilon,\varepsilon) & =\mathcal{C}(Q;\varepsilon+i0^{+},\varepsilon-i0^{+})=\frac{n_{i}^{2}u_{0}^{4}\int_{p}G^{R}(Q-p,\varepsilon)G^{A}(p,\varepsilon)}{1-n_{i}u_{0}^{2}\int_{p}G^{R}(Q-p,\varepsilon)G^{A}(p,\varepsilon)}=\frac{n_{i}u_{0}^{2}\zeta^{RA}(Q)}{1-\zeta^{RA}(Q)}\\
C^{RR}(Q;\varepsilon,\varepsilon) & C^{RR}(Q)=\mathcal{C}(Q;\varepsilon+i0^{+},\varepsilon+i0^{+})=\frac{n_{i}^{2}u_{0}^{4}\int_{p}G^{R}(Q-p,\varepsilon)G^{R}(p,\varepsilon)}{1-n_{i}u_{0}^{2}\int_{p}G^{R}(Q-p,\varepsilon)G^{R}(p,\varepsilon)}=\frac{n_{i}u_{0}^{2}\zeta^{RR}(Q)}{1-\zeta^{RR}(Q)}
\end{align}
At $T\to0$, $f'(\varepsilon)\to-\delta(\varepsilon)$, so only the
point $\varepsilon=0$ contributes. Suppressing $\varepsilon$ in
the arguments of $G$ and $C$,
\begin{equation}
\sigma^{\text{WL}}=2\pi\intop_{k,k^{\prime}}v_{k}v_{k'}\left[G^{A}(k)G^{A}(k')C^{RA}(Q)-G^{R}(k)G^{R}(k')C^{RR}(Q)\right]G^{R}(k')G^{R}(k)
\end{equation}
where $\varepsilon=0$ is understood and $Q=k+k'$.

To evaluate $C^{RA}(Q)$ and $C^{RR}(Q)$, we need to evaluate $\zeta^{RA}(Q)$
and $\zeta^{RR}(Q)$. To account for inelastic scattering which leads
to loss of phase coherence, we can phenomenologically modify $\zeta(Q)\to e^{-l/l_{\phi}}\zeta(Q)$,
where $l_{\phi}$ is the phase coherence length and $l$ is the mean
free path. Physically, this allows a probability $\propto e^{l/l_{\phi}}$
for the particle to lose phase coherence between successive elastic
scattering processes. 

For symmetric metals, the dominant -- in fact, divergent -- contribution
to $C^{RA}(Q)$ comes from $Q=0$ because $\zeta^{RA}(Q=0)$ turns
out to be 1 if $l/l_{\phi}\rightarrow0$. In contrast, the `RR' terms
are expected to be subdominant. 

\subsubsection{Momentum integrals}

After performing the frequency integral, we carry out the momentum
integral for a general asymmetric dispersion which results in an expression
involving the Fermi momenta of the two movers. We have
\begin{equation}
\zeta^{RA}(Q)=n_{i}u_{0}^{2}e^{-l/l_{\phi}}\int\frac{dp}{2\pi}\frac{1}{\epsilon_{Q-p}-\frac{i}{2\tau_{0}}}\frac{1}{\epsilon_{p}+\frac{i}{2\tau_{0}}}
\end{equation}
Now, there is no ``special'' value of $Q$ where $\epsilon_{p}=\epsilon_{Q-p}$
over all $p$. Nonetheless, the dominant contribution will presumably
come from terms where both $p$ and $Q-p$ are close to Fermi points,
$k_{F}^{R}$ and $-k_{F}^{L}$, so the corresponding $Q$ value is
$Q_{0}=k_{F}^{R}-k_{F}^{L}$. We have adopted a sign convention where
$k_{F}^{R},k_{F}^{L}>0$ are the magnitudes of the Fermi momenta.
Parametrizing (i) $p=k_{F}^{R}+P$, $Q-p=-k_{F}^{L}-P$, and (ii)
$p=-k_{F}^{L}+P$, $Q-p=k_{F}^{R}-P$ gives
\begin{align}
\zeta^{RA}(Q_{0}) & =\frac{n_{i}u_{0}^{2}e^{-l/l_{\phi}}}{\pi}\intop_{-\infty}^{\infty}dP\frac{1}{v_{L}P-\frac{i}{2\tau_{0}}}\frac{1}{v_{R}P+\frac{i}{2\tau_{0}}},
\end{align}
where we have extended the limits of the $P$-integral cutoffs to
$\pm\infty$ to focus on the contribution from regions near the Fermi
points. In practice, the cutoffs will be determined by the non-linearity
away from the Fermi points. This integral can be easily done by contour
methods and gives
\begin{align}
\zeta^{RA}(Q_{0}) & =e^{-l/l_{\phi}}\left(1-\frac{\delta v^{2}}{4v^{2}}\right)\label{eq:zeta-RA-Q0}
\end{align}
Clearly, $\zeta_{j}^{RA}\approx1$ for $\delta v\ll v$ and $l\ll l_{\phi}$.
Importantly, $C^{RA}(Q_{0})=\frac{n_{i}u_{0}^{2}\zeta^{RA}(Q_{0})}{1-\zeta^{RA}(Q_{0})}$
no longer diverges for $l/l_{\phi}\to0$ unlike the symmetric case. 

Now, to calculate a more accurate result, we need to consider nearby
momenta as well. For small deviations from $Q_{0}$, i.e., for $Q=Q_{0}+\Delta Q$,
the process can be repeated with (i) $p=k_{F}^{R}+P+\Delta Q/2$,
$Q-p=-k_{F}^{L}-P+\Delta Q/2$, and (ii) $p=-k_{F}^{L}+P+\Delta Q/2$,
$Q-p=k_{F}^{R}-P+\Delta Q/2$:
\begin{align}
\zeta^{RA}(Q_{0}+\Delta Q) & =\frac{\zeta^{RA}(Q_{0})}{1+\left[e^{l/l_{\phi}}\tau_{0}v\zeta^{RA}(Q_{0})\Delta Q\right]^{2}}\\
C^{RA}(Q_{0}+\Delta Q) & =\frac{\frac{v}{2\tau_{0}}\left[\zeta^{RA}(Q_{0})\right]^{2}}{\left[1-\zeta^{RA}(Q_{0})\right]+\left[e^{l/l_{\phi}}\tau_{0}v\zeta^{RA}(Q_{0})\Delta Q\right]^{2}}
\end{align}
where we have used Eq.~(\ref{eq:tau0}) with $\varepsilon=0$. Compared
to a symmetric dispersion which has $Q_{0}=0$, $\zeta^{RA}(Q_{0})=1$
and hence, a double pole in $C^{RA}$ at $Q=0$, $C^{RA}(Q_{0}+\Delta Q)$
has poles away from the real axis at 
\begin{equation}
\Delta Q=\pm i\frac{\sqrt{1-\zeta^{RA}(Q_{0})}}{e^{l/l_{\phi}}\tau_{0}v\zeta^{RA}(Q_{0})}\label{eq:CRA-poles}
\end{equation}

Similarly, 
\begin{equation}
\zeta^{RR}(Q_{0}+\Delta Q)=e^{-l/l_{\phi}}\frac{n_{i}u_{0}^{2}}{2\pi}2\text{Re}\intop_{-\infty}^{\infty}dP\left(\frac{1}{v_{L}(P-\Delta Q/2)+\frac{i}{2\tau_{0}}}\frac{1}{v_{R}(P+\Delta Q/2)+\frac{i}{2\tau_{0}}}\right)
\end{equation}
Now, both poles are above the real axis, so completing the contour
in the lower half-plane causes the integral to vanish exactly. Thus,
$\zeta^{RR}(Q_{0}+\Delta Q)=C^{RR}(Q_{0}+\Delta Q)=0$.

\subsection{Final result for $\sigma^{\text{WL}}$}

With the expression obtained after frequency and momentum integral,
we can calculate the correction to conductivity by doing another momentum
integral. Explicitly,
\begin{equation}
\sigma^{\text{WL}}=2\pi\intop_{k,k^{\prime}}v_{k}v_{k'}\frac{1}{\epsilon_{k}^{2}+\frac{1}{4\tau_{0}^{2}}}\frac{1}{\epsilon_{k'}^{2}+\frac{1}{4\tau_{0}^{2}}}C^{RA}(k+k')
\end{equation}
Again, we focus on pairs $(k,k')$ such that $k+k'=Q_{0}+\Delta Q$,
i.e., $k$ and $k'$ are near the left and right Fermi points or vice
versa. Parameterizing (i) $k=k_{F}^{R}+K+\Delta Q/2$, $k'=-k_{F}^{L}-K+\Delta Q/2$,
and (ii) $k=-k_{F}^{L}+K+\Delta Q/2$, $k'=k_{F}^{R}-K+\Delta Q/2$,
\begin{align}
\sigma^{\text{WL}} & =-2\pi\intop_{K,\Delta Q}v_{L}v_{R}\left[\frac{1}{v_{R}^{2}(K+\Delta Q/2)^{2}+\frac{1}{4\tau_{0}^{2}}}\frac{1}{v_{L}^{2}(K-\Delta Q/2)^{2}+\frac{1}{4\tau_{0}^{2}}}+\left(\Delta Q\to-\Delta Q\right)\right]C^{RA}(Q_{0}+\Delta Q)
\end{align}
For complex $K$, the above integral has simple poles at $K=\pm\frac{Q}{2}\pm\frac{i}{2v_{L,R}\tau_{0}}$.
Completing the $K$-contour in, say, the upper half plane gives
\begin{equation}
\sigma^{\text{WL}}=-2\pi\intop_{\Delta Q}\frac{4\tau_{0}^{3}ve^{l/l_{\phi}}\zeta^{RA}(Q_{0})}{1+\left[e^{l/l_{\phi}}\tau_{0}v\zeta^{RA}(Q_{0})\Delta Q\right]^{2}}C^{RA}(Q_{0}+\Delta Q)
\end{equation}
Besides the two simple poles of $C^{RA}(Q_{0}+\Delta Q)$ given by
Eq.\,(\ref{eq:CRA-poles}), we now have another pair of simple poles
at $\Delta Q=\pm i\left[e^{l/l_{\phi}}\tau_{0}v\zeta^{RA}(Q_{0})\right]^{-1}$.
Integrating over a complex half-plane gives
\begin{align}
\sigma^{\text{WL}} & =-\frac{2\pi\tau_{0}v\left[\zeta^{RA}(Q_{0})\right]^{2}}{\left[\sqrt{1-\zeta^{RA}(Q_{0})}+1-\zeta^{RA}(Q_{0})\right]}
\end{align}
 For large inelastic mean free path ($l\ll l_{\phi}$) and small asymmetry
($\left|\delta v\right|\ll v$), using Eq.\,(\ref{eq:zeta-RA-Q0})
and reinstating dimensionful factors of $e^{2}$ and $\hbar$ gives
Eq.\,(\ref{eq:sigma-WL}) in the main paper to leading orders in
$l/l_{\phi}$ and $\delta v/v$. 

\section{Recursive Green's function method\label{sec:Recursive-Green's-function}}

Recursive Green's function method is an iterative algorithm for calculating
properties of the system such as the localization length in the thermodynamic
limit \citep{RGF,RGF_Localization_Length,RGF_Original_Kramer_MacKinnon}.
As the name suggests, this method calculates the Green's function
recursively by using that of a smaller system size and growing the
system layer-by-layer. 

For a given Hamiltonian $H(N)$ for a system size of $N$, the Green's
function at complex energy $z$, $G(z,N)$ is defined as its resolvent
and for a real energy $E$, it is obtained by taking the imaginary
part of $z$ to zero. The localization length $(\xi)$ in terms of
the matrix elements of the Green's function is 
\begin{equation}
\frac{4}{\xi}=-\lim_{n\to\infty}\frac{1}{n}\ln\left(\text{Tr}|G_{1,n}|^{2}\right),\label{eq:RGF_Loc_Length}
\end{equation}
where $G_{n,m}\equiv\langle n|G(z,m)|m\rangle.$

The quantity $A_{n}=G_{1,n-2}^{-1}$, obeys the recursive relation
\citep{RGF_Localization_Length}:

\begin{equation}
A_{n+2}=(E-H_{n+1})V_{n}^{-1}A_{n+1}-V_{n}^{\dagger}V_{n-1}A_{n},\label{eq:RGF_Recursion}
\end{equation}
where $H_{n}$ is the matrix representing the tight-binding Hamiltonian
for the $n^{th}$ slice, and $V_{n}$ is the matrix that describes
the particles hopping onto the $(n+1)^{th}$ slice from the $n^{th}$
slice. For our model {[}Fig.\,\ref{fig:ToyModel_Dispersion}{]},
we have,

\begin{equation}
H_{n}=\left[\begin{array}{cc}
\epsilon_{i} & t\\
t & \epsilon_{i+1}
\end{array}\right],\,V_{n,n+1}=\left[\begin{array}{cc}
t'e^{-i\theta} & t\\
0 & t'e^{-i\theta}
\end{array}\right].\label{eq:RGF_Matrices_H_V}
\end{equation}

$\xi$ can be calculated by iterating Eq.\,(\ref{eq:RGF_Recursion})
with some initial values for $A_{0}$ and $A_{1}$, which we choose
as $A_{0}=0,A_{1}=V_{0}.$ However, Eq.\,(\ref{eq:RGF_Recursion})
suffers from a numerical instability in that the elements of $A_{n}$
grow exponentially for large $n$ and hence require some regularization.
Therefore, in every iteration we multiply both sides of Eq.\,(\ref{eq:RGF_Recursion})
with $[A_{n+1}]^{-1}$. Simplifying this procedure, we get the regularized
recursion relation \citep{RGF_Localization_Length},
\begin{equation}
\tilde{A}_{n}=(E-H_{n+1})V_{n}^{-1}-V_{n}^{\dagger}V_{n-1}^{-1}\tilde{A}_{n-1}^{-1},\label{eq:RGF_Recursion_Regularized}
\end{equation}
 that helps us resolve this issue and calculate $\xi$. We choose
$\tilde{A}_{0}=1$ and to calculate $\xi$, define a new matrix,

\begin{equation}
B_{n}=\frac{B_{n-1}\tilde{A}_{n}^{-1}}{b_{n}},
\end{equation}
where $b_{n}=||B_{n}||$ is the Frobenius norm of $B_{n}$, and $B_{0}=1.$
We calculate $B_{n}$ and store $b_{n}$ in every iteration of Eq.\,(\ref{eq:RGF_Recursion_Regularized}).
The matrix $B_{n}$ is very useful because

\begin{equation}
\ln\left(\text{Tr}|G_{1,n}|^{2}\right)=2\left[\ln(b_{n+1})+\cdot\cdot\cdot+\ln(b_{1})\right],\label{eq:RGF_Regularized_Trace}
\end{equation}
which can then be substituted in Eq.\,(\ref{eq:RGF_Loc_Length})
to determine $\xi$.\end{widetext}

\bibliographystyle{apsrev4-1}
\bibliography{library}

\end{document}